\documentclass[twocolumn,aps,showpacs,preprintnumbers,amsmath,amssymb, superscriptaddress,longbibliography]{revtex4-1}
\pdfoutput=1

\usepackage{bm}
\usepackage{graphicx}
\usepackage{color,xspace}

\renewcommand{\vec}{\mathbf}
\newcommand{\braket}[2]{\xspace\ensuremath{\langle #1\mid #2\rangle}\xspace}
\newcommand{\ket}[1]{\xspace\ensuremath{\mid #1\rangle}\xspace}

\begin{document}

\title{Observation of momentum-confined in-gap impurity state in Ba$_{0.6}$K$_{0.4}$Fe$_2$As$_2$: evidence for anti-phase $s_{\pm}$ pairing}

\author{P. Zhang}
\affiliation{Beijing National Laboratory for Condensed Matter Physics, and Institute of Physics, Chinese Academy of Sciences, Beijing 100190, China}
\author{P. Richard}\email{p.richard@iphy.ac.cn}
\affiliation{Beijing National Laboratory for Condensed Matter Physics, and Institute of Physics, Chinese Academy of Sciences, Beijing 100190, China}
\affiliation{Collaborative Innovation Center of Quantum Matter, Beijing, China}
\author{T. Qian}
\affiliation{Beijing National Laboratory for Condensed Matter Physics, and Institute of Physics, Chinese Academy of Sciences, Beijing 100190, China}
\author{X. Shi}
\affiliation{Beijing National Laboratory for Condensed Matter Physics, and Institute of Physics, Chinese Academy of Sciences, Beijing 100190, China}
\author{J. Ma}
\affiliation{Beijing National Laboratory for Condensed Matter Physics, and Institute of Physics, Chinese Academy of Sciences, Beijing 100190, China}
\author{L.-K. Zeng}
\affiliation{Beijing National Laboratory for Condensed Matter Physics, and Institute of Physics, Chinese Academy of Sciences, Beijing 100190, China}
\author{X.-P. Wang}
\affiliation{Beijing National Laboratory for Condensed Matter Physics, and Institute of Physics, Chinese Academy of Sciences, Beijing 100190, China}
\author{E. Rienks}
\affiliation{Helmholtz-Zentrum Berlin, BESSY, D-12489 Berlin, Germany}
\author{C.-L. Zhang}
\affiliation{Department of Physics and Astronomy, Rice University, Houston, Texas 77005, USA}
\affiliation{Department of Physics and Astronomy, The University of Tennessee, Knoxville, Tennessee 37996-1200, USA}
\author{Pengcheng Dai}
\affiliation{Beijing National Laboratory for Condensed Matter Physics, and Institute of Physics, Chinese Academy of Sciences, Beijing 100190, China}
\affiliation{Department of Physics and Astronomy, Rice University, Houston, Texas 77005, USA}
\author{Y.-Z. You}
\affiliation{Institute for Advanced Study, Tsinguha University, Beijing 100084, China}
\author{Z.-Y. Weng}
\affiliation{Institute for Advanced Study, Tsinguha University, Beijing 100084, China}
\affiliation{Collaborative Innovation Center of Quantum Matter, Beijing, China}
\author{X.-X. Wu}
\affiliation{Beijing National Laboratory for Condensed Matter Physics, and Institute of Physics, Chinese Academy of Sciences, Beijing 100190, China}
\author{J. P. Hu}
\affiliation{Beijing National Laboratory for Condensed Matter Physics, and Institute of Physics, Chinese Academy of Sciences, Beijing 100190, China}
\affiliation{Collaborative Innovation Center of Quantum Matter, Beijing, China}
\affiliation{Department of Physics, Purdue University, West Lafayette, Indiana 47907, USA}
\author{H. Ding}\email{dingh@iphy.ac.cn}
\affiliation{Beijing National Laboratory for Condensed Matter Physics, and Institute of Physics, Chinese Academy of Sciences, Beijing 100190, China}
\affiliation{Collaborative Innovation Center of Quantum Matter, Beijing, China}

\date{\today}

\begin{abstract}
We report the observation by angle-resolved photoemission spectroscopy of an impurity state located inside the superconducting gap of Ba$_{0.6}$K$_{0.4}$Fe$_2$As$_2$ and vanishing above the superconducting critical temperature, for which the spectral weight is confined in momentum space near the Fermi wave vector positions. We demonstrate, supported by theoretical simulations, that this in-gap state originates from weak scattering between bands with opposite sign of the superconducting gap phase. This weak scattering, likely due to off-plane non-magnetic Ba/K disorders, occurs mostly among neighboring Fermi surfaces, suggesting that the superconducting gap phase changes sign within holelike (and electronlike) bands. Our results impose severe restrictions on the models promoted to explain high-temperature superconductivity in these materials.\end{abstract}



\maketitle

When cooled below a critical temperature, superconducting (SC) materials develop an energy gap in their electronic structure that is a direct signature of the interactions leading to the condensation of pairs of charge carriers. This SC gap is a macroscopic property fully characterized by an amplitude and a phase. While there is sufficient experimental evidence for a Fermi surface dependent superconducting gap amplitude in the multi-band Fe-based high-temperature superconductors, the sign of the phase on their various Fermi surface sheets remains highly controversial, most popular proposals suggesting either a common phase throughout the momentum space \cite{Kontani_PRL104} or a phase with a sign alternating between holelike and electronlike Fermi surface sheets \cite{MazinPRL2008,KurokiPRL101,SeoPRL2008}. The knowledge of the relative sign of the phase on the various energy bands forming their Fermi surface (FS) would impose severe restrictions on the validity of the numerous models invoked to explain high-temperature superconductivity in these systems. In addition, some controversy remains on the amplitude of the SC gap at the Brillouin zone center ($\Gamma$). Although most ARPES reports indicate a strong coupling for the inner hole pockets in optimally-doped Ba$_{1-x}$K$_{x}$Fe$_2$As$_2$ \cite{Ding_EPL, L_Zhao_CPL25,Wray_PRB2008,Evtushinsky_PRB89}, sub-BCS coupling has been reported in a laser-ARPES study \cite{Shimojima_Science332}. 

In this paper we study how the low-energy electronic states couple with impurities in the SC state as a tool to infer the relative sign of the SC gap phase on the FS sheets of the optimally-doped Ba$_{0.6}$K$_{0.4}$Fe$_2$As$_2$ ferropnictide. Our angle-resolved photoemission spectroscopy measurements clarify the amplitude of the SC gap on the $\Gamma$-centered hole pockets and indicate an in-gap state located around 6 meV. Using theoretical simulations and assuming that the relevant impurities are non-magnetic, we show that this state arises from scattering between bands with opposite sign of the SC gap phase. Our results are consistent with the orbital anti-phase $s_{\pm}$ gap structure and impose severe restrictions to the models promoted to explained high-temperature superconductivity in the Fe-based superconductors. 

Large single-crystals of Ba$_{0.6}$K$_{0.4}$Fe$_2$As$_2$ with high quality were grown using the self-flux method \cite{GF_ChenPRB78}, and their $T_c$ was determined to be 37 K from magnetization measurements. ARPES measurements were performed at the 1-cubed ARPES end-station of BESSY and in our own facility at the Institute of Physics, Chinese Academy of Sciences, using a VG-Scienta R4000 electron analyzer and 21.2 eV photons. The light used in BESSY was linearly polarized in directions parallel or perpendicular to the analyzer slit. The angular resolution was set to 0.2$^{\circ}$ whereas the energy resolution to 4-5 meV in BESSY and to 3-4 meV at the Institute of Physics. Clean surfaces for the ARPES measurements were obtained by cleaving the samples \emph{in situ} in a working vacuum better than 5 $\times$ 10$^{-11}$ Torr. In the text we label the momentum values with respect to the 1 Fe/unit cell BZ.

\begin{figure*}[!t]
\begin{center}
\includegraphics[width=\textwidth]{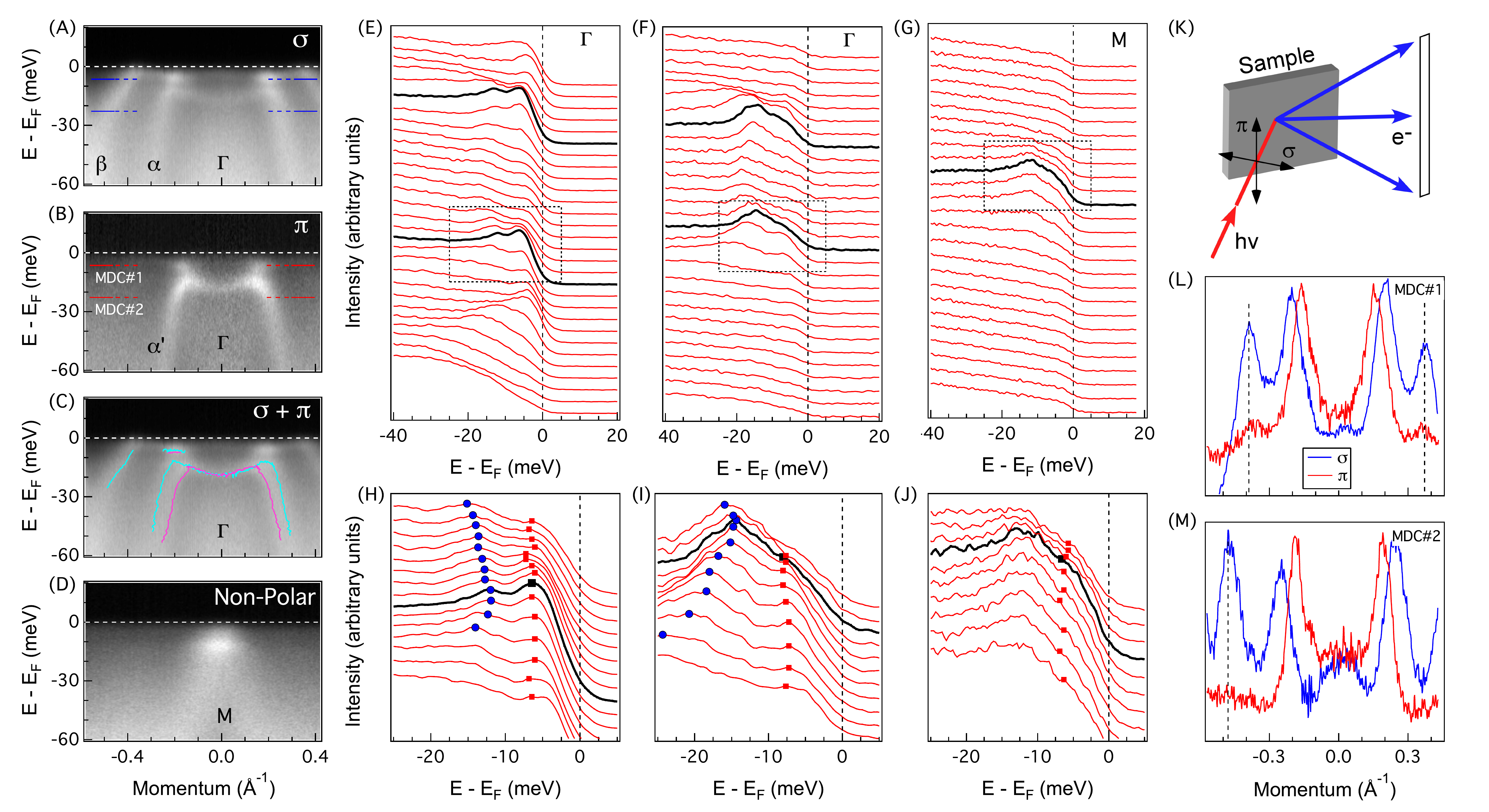}
\end{center}
 \caption{\label{vs_polarization} (\emph{A-B}) ARPES intensity plots for a cut passing through $\Gamma$ and oriented along $\Gamma$-M recorded at 1 K using $\sigma$ and $\pi$ incident light polarizations, respectively. The blue and red dashed lines indicate the position of the MDCs in panels \emph{L} and \emph{M}. (\emph{C}) Sum of the ARPES intensity plots in \emph{A} and \emph{B}. The turquoise and pink curves represent the band dispersions extracted from the MDCs and EDCs in \emph{A} and \emph{B}, respectively. (\emph{D}) ARPES intensity plot for a cut passing near the M point, measured at 4.2 K with unpolarized light. (\emph{E-G}) EDC plots corresponding to the cuts shown in panels \emph{A}, \emph{B} and \emph{D}, respectively. The EDCs in black correspond to the $k_F$ positions. (\emph{H-J}) Near-$E_F$ zooms corresponding the the dashed boxes in panels \emph{E}, \emph{F} and \emph{G}, respectively. Circles and squares are used to identify the $\alpha/\alpha'$ band and the impurity state, respectively, as obtained by the maximum positions in the EDCs. (\emph{K}) Schematic representation of the polarization configurations. (\emph{L}) Comparison of the MDCs recorded with $\sigma$ and $\pi$ polarizations at the MDC\#1 position (7 meV below $E_F$) in panels \emph{A} (blue) and \emph{B} (red). The vertical dashed lines are guides to the eye for the peak positions. (\emph{M}) Same as \emph{L} but for the MDC\#2 position (22 meV below $E_F$).}
\end{figure*}

In Fig. \ref{vs_polarization}\emph{A} we display the angle-resolved photoemission spectroscopy (ARPES) intensity plot of a cut passing through the Brillouin zone (BZ) center of optimally-doped Ba$_{0.6}$K$_{0.4}$Fe$_2$As$_2$ ($T_c=37$ K) and oriented along the $\Gamma (0, 0)$-M$(\pi,0)$ direction, which has been recorded at 1 K using $\sigma$ incident light polarization, as defined in Fig. \ref{vs_polarization}\emph{K}. This configuration is sensitive to electronic states with odd symmetry, such as the $\alpha$ band formed mainly by the odd combination of the $d_{xz}$ and $d_{yz}$ orbitals, and the $\beta$ band associated with the $d_{xy}$ orbital \cite{XP_WangPRB85}. In agreement with previous ARPES results \cite{Ding_EPL, L_Zhao_CPL25}, the corresponding energy distribution curve (EDC) plot in Fig. \ref{vs_polarization}\emph{E} and the corresponding near-$E_F$ zoom shown in Fig. \ref{vs_polarization}\emph{H} reveal a SC gap of about 12 meV on the $\alpha$ band and of about 6 meV on the $\beta$ band. Interestingly, the EDC at the Fermi wave vector ($k_F$) location of the $\alpha$ band exhibits a strong peak at 6 meV, inside the SC gap, which has been reported already in an early ARPES report \cite{Ding_EPL}. Similar observation is made for the cut displayed in Fig. \ref{vs_polarization}\emph{B}, which has been recorded in the same conditions but with $\pi$ polarization, which is sensitive to the $\alpha$' band formed by the even combination of the $d_{xz}$ and $d_{yz}$ orbitals \cite{XP_WangPRB85}. As illustrated by the corresponding EDC plot shown in Fig. \ref{vs_polarization}\emph{F} and the near-$E_F$ zoom displayed in Fig. \ref{vs_polarization}\emph{I}, the $\alpha$' band is gapped by 13 meV and a peak inside the SC gap is also detected at 7 meV.

In Fig. \ref{vs_polarization}\emph{C}, we show the ARPES intensity plot obtained by summing the intensity plots recorded with $\sigma$ and $\pi$ polarizations, and we overlay the electronic band dispersions obtained by tracking the various peak positions in the EDCs and the momentum distribution curves (MDCs). At this particular photon energy (21.2 eV), the $\alpha$ and $\alpha$' bands are very close in momentum but the use of variable polarization allows to separate them. More importantly, when the energy reaches the minimum gap location, they both show a bending back characteristic of the Bogoliubov dispersion in the SC state, confirming the SC origin of the peaks observed at 12-13 meV on the $\alpha$ and $\alpha$' bands. In addition, our analysis allows us to follow the dispersionless in-gap feature over a short momentum range around the $k_F$ positions of the $\alpha$ and $\alpha$' bands. By comparing the MDCs recorded at 7 meV and 22 meV below the Fermi energy ($E_F$) in Figs. \ref{vs_polarization}\emph{L} and \ref{vs_polarization}\emph{M}, we can see that the in-gap peak is also observed at the $k_F$ position of the $\beta$ band. A peak is detected for the MDC recorded 7 meV below $E_F$ in the $\pi$ polarization configuration (red curve in Fig. \ref{vs_polarization}\emph{L}), whereas the $\beta$ band is completely suppressed, as suggested by the MDC recorded 22 meV below $E_F$ (red curve in Fig. \ref{vs_polarization}\emph{M}). The observation of the in-gap state in both odd and even configurations of polarization contrasts with the expectation for a regular energy band carrying a specific orbital character, and rather suggests a mixture of several orbital characters. Interestingly, a similar in-gap state is also observed around the M point, as shown in the ARPES intensity plot displayed in Fig. \ref{vs_polarization}\emph{D} and the corresponding EDC plots shown in Figs. \ref{vs_polarization}\emph{G} and \ref{vs_polarization}\emph{J}.

\begin{figure}[!t]
\begin{center}
\includegraphics[width=0.48\textwidth]{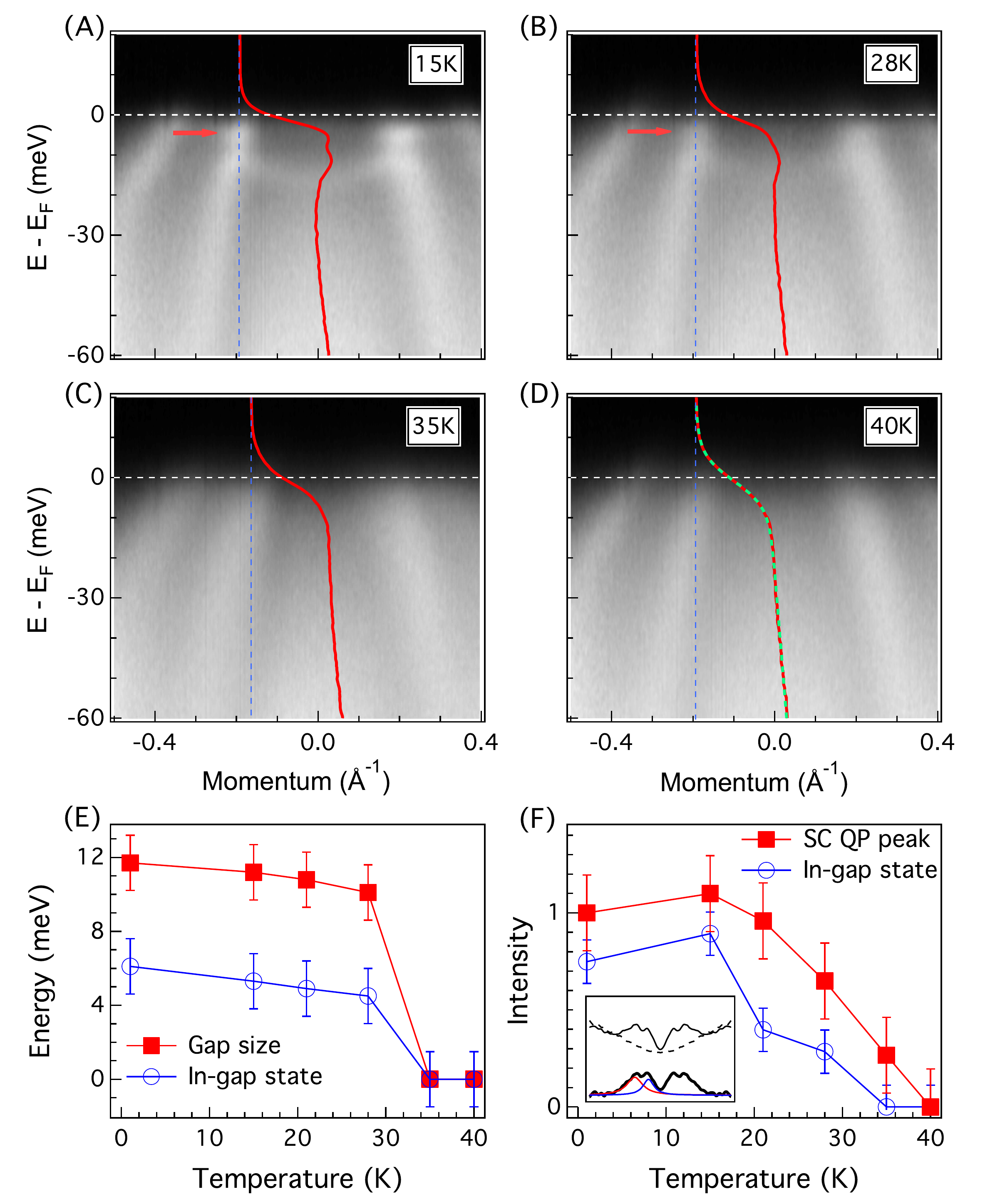}
\end{center}
 \caption{\label{vs_Temperature} (\emph{A-D}) Temperature evolution, from (\emph{A}) 15 K to (\emph{D}) 40 K, of the ARPES intensity plots for the cut presented in Fig. \ref{vs_polarization}\emph{A}. The red arrows indicate the in-gap states. The EDCs at $k_F$ are also displayed in red. The EDC in \emph{D} does not show any in-gap state, as suggested by a fit (in green) to the Fermi-Dirac function convoluted with the resolution function. (\emph{E}) Temperature evolution of the SC gap amplitude (red) and the position of the in-gap state peak, as determined from EDC analysis. (\emph{F}) Temperature dependence of the spectral weight of the SC coherent peak (red) and in-gap state peak at $k_F$, as determined from EDC analysis. The intensity is normalized to the area under the fit curve of the SC coherent peak at 1 K. The inset illustrates how the spectral weight is extracted. After symmetrizing the EDCs to approximately remove the Fermi cutoff, and removing a background shown by a dashed line, we fit the left part of the subtracted symmetrized EDC using two Lorentzian curves. The spectral weight is given by the area below each Lorentzian curve.}
\end{figure}

In order to discuss further the origin of the in-gap state, we display in Figs. \ref{vs_Temperature}\emph{A}-\ref{vs_Temperature}\emph{D} ARPES intensity plots recorded between 15 K and 40 K. Surprisingly, the in-gap state is clearly visible up to 28 K, but is undetectable near and above $T_c$. Our quantitative analysis, presented in Figs. \ref{vs_Temperature}\emph{E} and \ref{vs_Temperature}\emph{F}, indicates that similarly to the SC gap, the binding energy position of the in-gap state decreases only slightly as temperature increases, and that its peak intensity follows roughly the intensity of the SC coherent peak, thus suggesting a close relationship between the in-gap state and superconductivity. In fact, a previous study based on laser-ARPES associated the in-gap state peak to a SC gap, while assigning the peak at 12 meV to a magnetic resonance mode or a coupling with orbital degrees of freedom \cite{Shimojima_Science332}. We argue here that this scenario is incompatible with the flatness of the in-gap state dispersion, with the finite weight of the in-gap state near the $\beta$ band, as well as with the obvious Bogoliubov dispersion for the features around 12-13 meV, which disappear at the bulk $T_c$ value. 

\begin{figure}[!t]
\begin{center}
\includegraphics[width=0.48\textwidth]{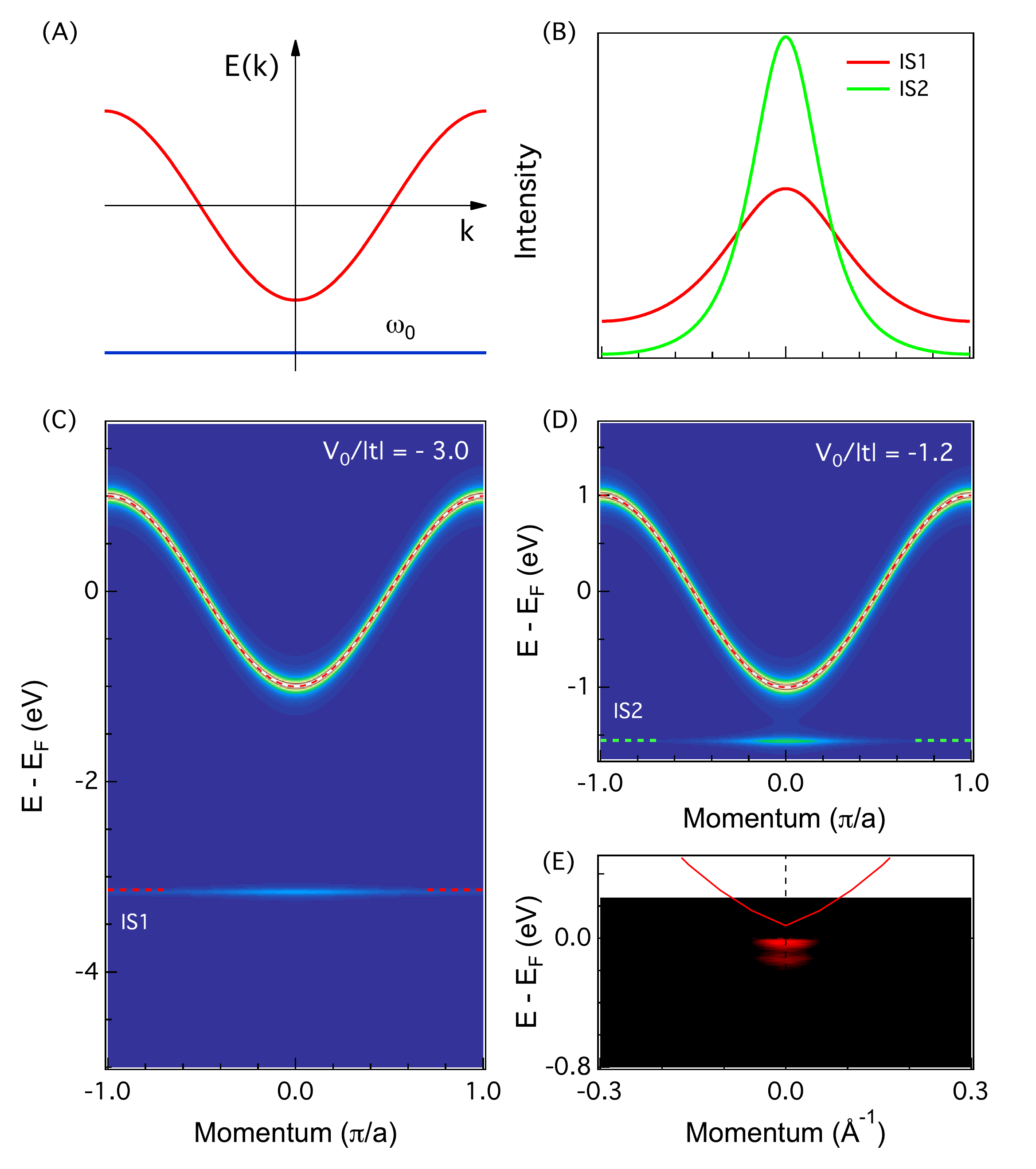}
\end{center}
 \caption{\label{Fig_Ga2O3} (\emph{A}) Schematic illustration of an impurity level below a dispersive band. (\emph{B}) Comparison of the MDC at the impurity level from panels \emph{C} and \emph{D}. (\emph{C}-\emph{D}) Numerical simulations of the spectral weight for a system with a dispersive band of the form $\varepsilon(k)=t\cos(\pi k)$ (with $t=-1$ eV) scattering on 5\% of impurities according to Eq. [\ref{one_band}], with $V_0/|t|=-3.0$ and $V_0/|t|=-1.2$, respectively. (\emph{E}) Impurity state in Si-doped $\beta$-Ga$_2$O$_3$, from Ref. \cite{RichardAPL2012}. The red curve represents the band structure obtained from LDA calculations.}
\end{figure}

Because its intensity is sample dependent, an early ARPES study pointed out that the in-gap state could be related to impurities \cite{Ding_EPL}. One interesting characteristic of the in-gap state is that its spectral weight is mainly confined to small momentum regions near the $k_F$ positions of the bands. In principle, an impurity state well localized in real space is expected to extend all over the momentum space, as illustrated in Fig. \ref{Fig_Ga2O3}\emph{A}. However, momentum space localization of the spectral weight is possible when the scattering is described by a weak potential. In Fig. \ref{Fig_Ga2O3}\emph{C}, we show the result of a simulation in which a single non-magnetic impurity per N sites is introduced in a system and scatters weakly, as a $\delta(\vec{r})$ potential, with an electronic dispersive band of the form $\varepsilon(k)=t\cos(\pi k)$. More precisely, we describe the system by the Hamiltonian \cite{BalatskyRMP}:

\begin{equation}\label{one_band}
H=\sum_{\vec{k},\sigma}\varepsilon_{\vec{k},\sigma}c^{\dagger}_{\vec{k},\sigma}c_{\vec{k},\sigma}+\frac{V_0}{N}\sum_{{\vec{k}},{\vec{k'}},\sigma}c^{\dagger}_{\vec{k},\sigma}c_{\vec{k'},\sigma},
\end{equation}

\noindent where the first term represents the unperturbed Hamiltonian with the operator $c^{\dagger}_{\vec{k},\sigma}$($c_{\vec{k},\sigma}$) creating(anihilating) an electron of spin $\sigma$ and wave vector $\vec{k}$, while the second term is the perturbation, with the potential $V_0$ characterizing the strength of the interaction. We divided the first BZ into 500 points and diagonalized the resulting matrices in eq. \eqref{one_band} numerically. The spectral function $A(\vec{k},\omega)$ was extracted by using the equation:

\begin{equation}
\label{spectral_weight}
A(\vec{k},\omega)=-\frac{1}{\pi}\displaystyle\textrm{Im}\sum_m\frac{|\braket{\vec{k}}{m}|^2}{\omega-E_m+i\delta},
\end{equation} 

\noindent where the eigenvectors $\ket{m}$ with eigenvalues $E_m$ are projected into the momentum space. We note that only the diagonal terms of the Green's function are used to illustrate the resulting spectral weight in Fig. \ref{Fig_Ga2O3}\emph{C}.

As expected, our calculations reveal a dispersionless state, which is located below the dispersive band, in the band gap. Interestingly though, the spectral weight of the impurity state is not evenly distributed throughout the momentum space, but is larger near the bottom of the electronic band. This is a manifestation that the impurity state is not completely localized in real space and that it hybridizes with the dispersive band. In Fig. \ref{Fig_Ga2O3}\emph{D}, we show the results of the calculations for a weaker impurity potential $V_0$ ($-1.2|t|$ as compared to $-3|t|$). As the impurity potential decreases, the impurity state locates closer to the bottom of the dispersive band, and its spectral weight becomes more localized in the momentum space. 

The latter observation is well illustrated by the MDCs at the impurity levels, which are compared in Fig. \ref{Fig_Ga2O3}\emph{B}. As illustrated in Fig. \ref{Fig_Ga2O3}\emph{E}, the formation of such in-gap impurity states with spectral weight limited in momentum space has been observed experimentally \cite{RichardAPL2012} in Si-doped $\beta$-Ga$_2$O$_3$, a large-gap semiconductor, and attributed to the semi-localization of the impurity states consistent with the Heisenberg uncertainty principle $\Delta k\Delta x\sim 1$, in agreement with our simulations and with scanning tunneling spectroscopy measurements on the same material \cite{IwayaAPL2011}. 

\begin{figure*}[!t]
\begin{center}
\includegraphics[width=\textwidth]{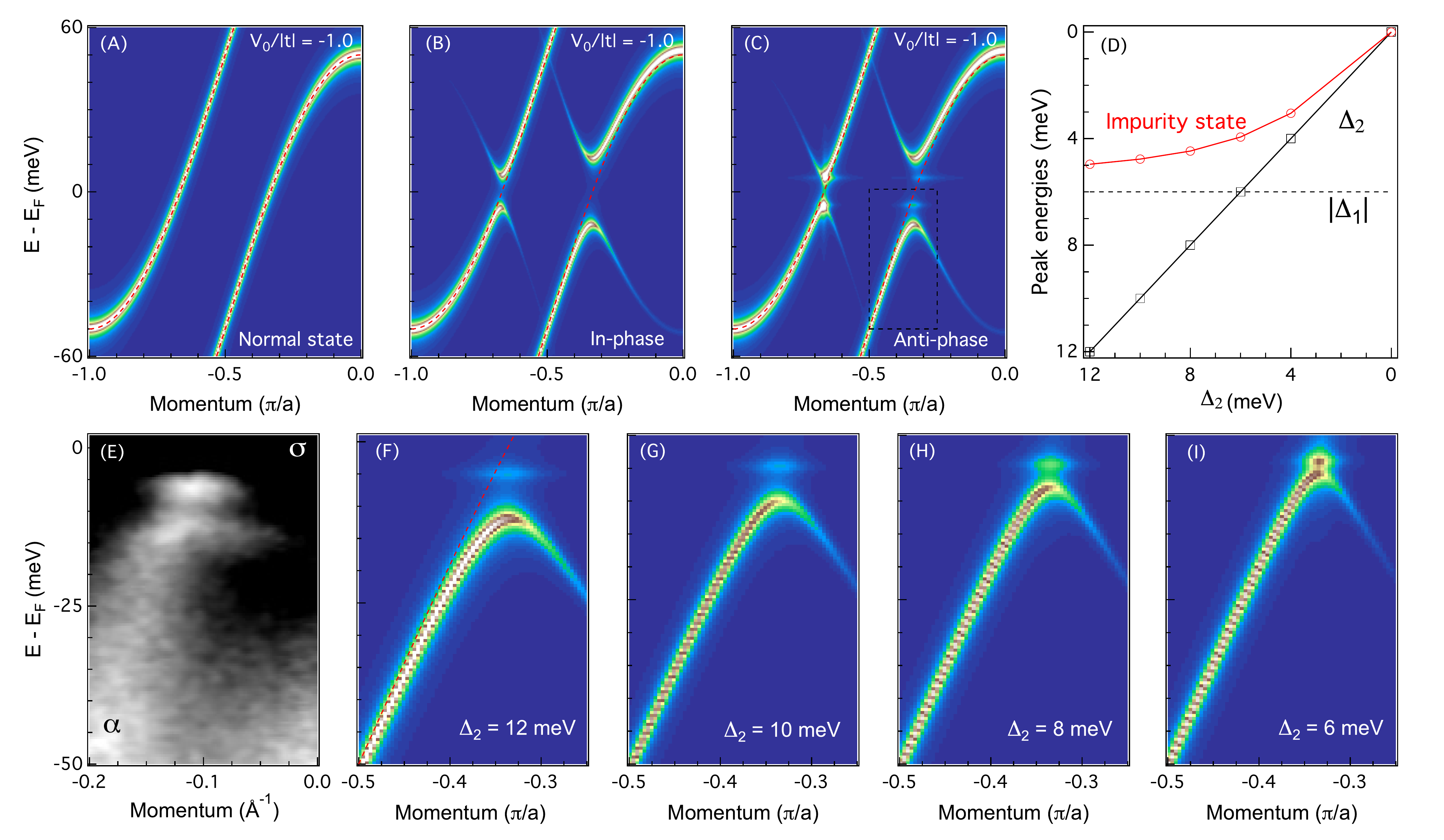} 
\end{center}
 \caption{\label{Fig_calculations} (\emph{A}-\emph{C}) Numerical simulations of the spectral weight for a system with an impurity interacting with two dispersive bands according to Eq. [\ref{two_bands}] in the normal state ($\Delta_1=\Delta_2=0$), in the in-phase SC state ($\Delta_1=\frac{1}{2}\Delta_2\neq 0$), and in the anti-phase SC state ($\Delta_1=-\frac{1}{2}\Delta_2\neq 0$), respectively. In all cases, we used $V_0/|t|=-1$. The red dashed lines represent the electronic dispersion in the normal state. (\emph{D}) Simulation of the energy position of the in-gap impurity state as a function of $|\Delta_2|$ for a fixed $|\Delta_1|$. (\emph{E}) Zoom on the impurity state found experimentally near the $\alpha$ band of Ba$_{0.6}$K$_{0.4}$Fe$_2$As$_2$. (\emph{F}-\emph{I}) Numerical simulation of the zoom near the impurity state corresponding to the dashed box from panel \emph{C} (anti-phase SC), as a function of $|\Delta_2|$, for $|\Delta_1|=6$ meV (anti-phase gaps). The red dashed line in \emph{F} represents the bare band dispersion.}
\end{figure*}

We now ask: how would in-gap impurity states manifest themselves in a superconductor? Early after the discovery of the Fe-based superconductors and the proposal of $s_{\pm}$ pairing symmetry \cite{MazinPRL2008,KurokiPRL101,SeoPRL2008}, a few theoretical studies pointed out that impurities could affect the density-of-states inside the SC gap \cite{ParishPRB78,ParkerPRB78,ChubukovPRB78,BangPRB78}, a proposal that gained in popularity recently due to its possible ability to distinguish the $s_{\pm}$ and $s_{++}$ gap structures of the Fe-based superconductors \cite{Y_WangPRB87}. Rather than limiting ourselves to the density-of-states, here we investigate this problem from the point of view of a momentum-resolved probe. Since it is well-known that there should be no bounded in-gap state in the case of a one-band $s$-wave system with non-magnetic impurities \cite{PW_AndersonJPCS11}, we performed simulations on a two-band system with a non-magnetic impurity, a choice first justified by the lack in the literature of observation of magnetic impurities in as-grown (Ba,K)Fe$_2$As$_2$ and further explained below. We thus used the Hamiltonian:

\begin{eqnarray}
H&=&\sum_{\vec{k},m,\sigma}\varepsilon_m(\vec{k})c^{\dagger}_{m,\vec{k},\sigma}c_{m,\vec{k},\sigma} \notag\\
&+&\sum_{\vec{k},m}\Delta_m(c^{\dagger}_{m,\vec{k},\uparrow}c^{\dagger}_{m,-\vec{k},\downarrow}+c_{m,\vec{k},\downarrow}c_{m,-\vec{k},\uparrow}) \notag\\
&+&\frac{V_0}{2N}\sum_{m,n,\vec{k},\vec{k'},\sigma}c^{\dagger}_{m,\vec{k},\sigma}c_{n,\vec{k'},\sigma},\label{two_bands}
\end{eqnarray}

\noindent where the second term accounts for the SC gap and with the indexes $m$ and $n$ taking the values 1 and 2 representing the two bands. Fig. \ref{Fig_calculations}\emph{A} shows the result in the normal state, \emph{i. e.} for null SC gaps ($\Delta_1=\Delta_2=0$). No impurity state is observed near $E_F$ in this case. We then consider two cases for which a SC gap opens on the two bands, with a 2:1 size ratio. The relative phase of the gap on the two bands is given directly by the sign of the parameters $\Delta_{1,2}$. Fig. \ref{Fig_calculations}\emph{B} displays the result for in-phase SC gaps ($\Delta_1=\frac{1}{2}\Delta_2\neq 0$). In this condition, no impurity state is observed. In contrast, our results for the anti-phase case ($\Delta_1=-\frac{1}{2}\Delta_2\neq 0$), shown in Fig. \ref{Fig_calculations}\emph{C}, reveal an in-gap state with a spectral weight confined in momentum space around the $k_F$ position of the two bands, as well illustrated by the zoom in Fig. \ref{Fig_calculations}\emph{F}. This phenomenon is very similar to our experimental observation on Ba$_{0.6}$K$_{0.4}$Fe$_2$As$_2$, for which a near-$E_F$ zoom at the $k_F$ position of the $\alpha$ band is displayed in Fig. \ref{Fig_calculations}\emph{E}. Although quantitative differences are found experimentally between two-band and five-band models \cite{Gastiasoro_PRB89}, our simulations certainly provide an illustration of the physics at the origin of the in-gap impurity state observed in our study. 

The doping evolution of the low-energy electronic structure in the SC state of Ba$_{1-x}$K$_{x}$Fe$_2$As$_2$ has been shown in a laser-ARPES study \cite{Malaeb_PRB86}. Within the context of the current interpretation of the low-energy feature as an in-gap state, these data provide some indications on the behavior of the in-gap state as a function of doping. With doping increasing, the energy position of the in-gap state peak decreases only slightly, and is no longer resolved for $x\geq 0.6$, doping at which the high-energy feature here interpreted as the SC gap, decreases to reach the same magnitude. Interestingly, this effect is easily reproduced by our simulations, presented in Figs. \ref{Fig_calculations}\emph{F}-\ref{Fig_calculations}\emph{I}, in which we vary the size of the gap $|\Delta_2|$ while maintaining $|\Delta_1|=6$ meV constant. Indeed, the position of the in-gap state decreases only slightly with $|\Delta_2|$ decreasing, and the two features are very close in energy for  
$|\Delta_2|=6$ meV, as if the two features were merging. This effect is summarized in Fig. \ref{Fig_calculations}\emph{D}, where $|\Delta_2|$ is decreased down to 0. Although the experimental case is necessarily more complicated since more bands are involved and all the gaps evolve with doping, our simulations reinforce our interpretation of the 6 meV peak as an in-gap impurity state and calls for a reinterpretation of the laser-ARPES data given in Ref. \cite{Malaeb_PRB86}.

As a corollary from our simulations, our observation of an in-gap state implies directly that if the related impurities are non-magnetic, the SC gaps on the various FS pockets do not have all the same phase, which discards all scenarios favoring a $s_{++}$ pairing symmetry, such as the low-energy orbital fluctuation model \cite{Kontani_PRL104}. To derive further consequences of our observation, we now try to determine the origin of the scattering giving rise to the impurity state. Instead of the constant potential $V_0$, from here we consider the screened Coulomb potential:

\begin{equation}
\label{screening_r}V(r)=-\frac{e^2}{4\pi\epsilon_0 \sqrt{r^2+d^2}}e^{-\sqrt{r^2+d^2}/\lambda},
\end{equation}

\noindent where $d$ is the distance between the Fe and impurity planes, $r$ is the in-plane distance and $\lambda$ is the Thomas-Fermi screening length, which is estimated to be about 1 \AA\xspace for the $1.85\times 10^{21}$ cm$^{-3}$ electron density in Ba$_{0.55}$K$_{0.45}$Fe$_2$As$_2$ \cite{JohnstonAdv_Phys2010}. We first consider the trivial case where the K$^{+}$ ions doping the Ba$_{0.6}$K$_{0.4}$Fe$_2$As$_2$ system at the Ba$^{2+}$ site act as non-magnetic weak potential centers for the in-plane Fe electronic states. Using $d= 3.1$ \AA\xspace for the distance from the dopant atom to the Fe plane, we can deduce that the scattering potential at the nearest Fe site is about 0.24 eV. This potential is slightly smaller than the 0.3 eV kinetic energy estimated from the 0.6 eV bandwidth of the $\alpha$ band in Ba$_{0.6}$K$_{0.4}$Fe$_2$As$_2$ \cite{Ding_JPCM2011}. In contrast, assuming $d=0$ for an in-plane impurity, the largest $V(r)$ potential would be reached for a distance $r$ comparable to a $\sim 1$ \AA\xspace screening length. This situation leads to a scattering potential of about 5.3 eV, which cannot be regarded as weak. Consequently, even if a magnetic in-plane impurity were present, such as a Fe vacancy, our simple calculation suggests that it would not produce the momentum-confined feature observed here in Ba$_{0.6}$K$_{0.4}$Fe$_2$As$_2$. Therefore, the ARPES signature of a momentum-confined spectral weight for the impurity state in Ba$_{0.6}$K$_{0.4}$Fe$_2$As$_2$ is more consistent with non-magnetic out-of-plane scattering by the Ba$^{2+}$/K$^+$ disorder. This may also explain why the signature of impurity scattering is not common to all Fe-based superconductors. Indeed, many ferropnictides and ferrochalcogenides are doped directly in the FeAs planes, which leads to strong scattering preventing the observation of well-defined momentum confined impurity states. Despite the very clear superconducting features it exhibits in the ARPES spectra \cite{UmezawaPRL2012,BorisenkoSymmetry4}, LiFeAs is nominally impurity free and thus no impurity state is found experimentally. 

\begin{figure}[!t]
\begin{center}
\includegraphics[width=0.48\textwidth]{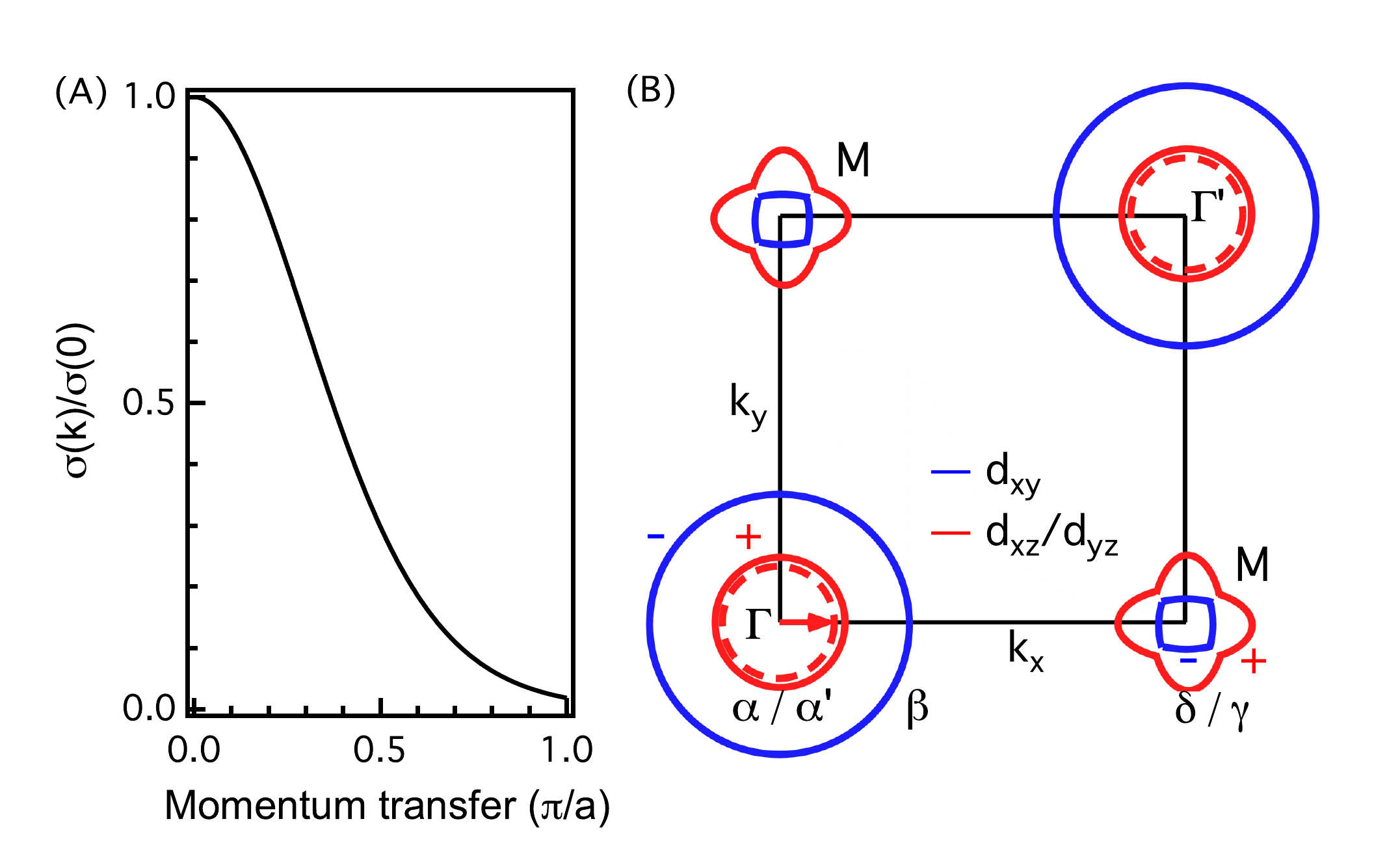} 
\end{center}
 \caption{\label{Fig_scattering} (\emph{A}) Calculated scattering strength as a function of momentum transfer (see the text). (\emph{B}) Schematic FS of Ba$_{0.6}$K$_{0.4}$Fe$_2$As$_2$, with the FS sheets drawn in red and blue having opposite SC gap phase sign.}
\end{figure}

In Fig. \ref{Fig_scattering}(A), we display the scattering strength as function of the momentum transfer, obtained by calculating the Fourier transform of Eq. [\ref{screening_r}]. This strength decreases smoothly as a function of the momentum transfer. For a momentum transfer corresponding to the $\Gamma$-M distance, the scattering strength is already 54 times smaller than the strength for 0 momentum transfer scattering. Although impurity scattering-induced momentum transfer is possible between any points in the BZ, this suggests that small momentum transfer, \emph{i. e.} forward scattering, is favored. 

Our experiment does not allow us to determine precisely the relationship between the sign of the phase on all the bands. Nevertheless, the latter argument, namely that forward scattering is favored, suggests that: i) one of the three $\Gamma$-centered hole FS pocket must have a different SC gap phase sign from the two others, and ii) the two M-centered electron FS pockets must be associated with opposite sign of the SC gap phase. This situation, illustrated in Fig. \ref{Fig_scattering}\emph{B} contrasts with the conventional $s_{\pm}$ pairing symmetry, for which the sign of the gap on the three hole FS pockets is identical and differs from that on both of the electron FS pockets. 

A previous study on the four-site model has shown that sign change exists between two hole pockets because the $d_{xy}$ orbital has a sign opposite to the $d_{yz}$ and $d_{xz}$ orbitals \cite{X_LuPRB85}. A recent study using first-principle calculations, including the \emph{ab initio} determination of the two-particle vertex function, reveals the existence of three competing gap configurations \cite{Yin_ZP_orbital2013} and also suggests a similar orbital dependent sign change. However, in general, this state should have nodes on the electron pockets if the sign change between them exists or shows large anisotropicity of the SC gaps. Although a non-negligible anisotropicity of the SC gap was observed by ARPES in LiFeAs \cite{UmezawaPRL2012,BorisenkoSymmetry4}, the anisotropicities in most Fe-based superconductors are small \cite{YB_HuangAIP_Adv2}. Most recently, it was also suggested that an odd-parity SC pairing term \cite{HuJP_PRX3} can produce the exact anti-phase $s_{\pm}$ state \cite{Hao_PRB89,HuJP_PRX2012}. In the absence of hole pockets, the odd parity pairing is the same as the bonding-antibonding $s_{\pm}$ suggested in Refs. \cite{MazinPRB84,KhodasPRL108}. However, such proposal remains to be tested experimentally.

Interestingly, the sign change within the holelike FS pockets and within the electron like pockets characterizing the anti-phase $s_{\pm}$ state is consistent with many experimental results reported previously for both ferropictides and ferrochalcogenides \cite{Hao_PRB89}. Therefore, although more experiments are required in the future, we can speculate that the anti-phase $s_{\pm}$ may be the state unifying all families of Fe-based superconductors. 

We acknowledge X. Dai, G. Kotliar and Z.-Q. Wang for useful discussions and thank N. Xu and Y.-B. Huang for technical assistance. This work was supported by grants from CAS (2010Y1JB6), MOST (2010CB923000,  2011CBA001000, 2011CBA00102, 2012CB821403 and 2013CB921703) and NSFC (11004232, 11034011/A0402, 11234014 and 11274362) from China. The single crystal growth at Rice/UTK was supported by the US DOE, BES, through contract DE-FG02-05ER46202

\bibliography{biblio_long}

\begin{thebibliography}{34}%
\makeatletter
\providecommand \@ifxundefined [1]{%
 \@ifx{#1\undefined}
}%
\providecommand \@ifnum [1]{%
 \ifnum #1\expandafter \@firstoftwo
 \else \expandafter \@secondoftwo
 \fi
}%
\providecommand \@ifx [1]{%
 \ifx #1\expandafter \@firstoftwo
 \else \expandafter \@secondoftwo
 \fi
}%
\providecommand \natexlab [1]{#1}%
\providecommand \enquote  [1]{``#1''}%
\providecommand \bibnamefont  [1]{#1}%
\providecommand \bibfnamefont [1]{#1}%
\providecommand \citenamefont [1]{#1}%
\providecommand \href@noop [0]{\@secondoftwo}%
\providecommand \href [0]{\begingroup \@sanitize@url \@href}%
\providecommand \@href[1]{\@@startlink{#1}\@@href}%
\providecommand \@@href[1]{\endgroup#1\@@endlink}%
\providecommand \@sanitize@url [0]{\catcode `\\12\catcode `\$12\catcode
  `\&12\catcode `\#12\catcode `\^12\catcode `\_12\catcode `\%12\relax}%
\providecommand \@@startlink[1]{}%
\providecommand \@@endlink[0]{}%
\providecommand \url  [0]{\begingroup\@sanitize@url \@url }%
\providecommand \@url [1]{\endgroup\@href {#1}{\urlprefix }}%
\providecommand \urlprefix  [0]{URL }%
\providecommand \Eprint [0]{\href }%
\providecommand \doibase [0]{http://dx.doi.org/}%
\providecommand \selectlanguage [0]{\@gobble}%
\providecommand \bibinfo  [0]{\@secondoftwo}%
\providecommand \bibfield  [0]{\@secondoftwo}%
\providecommand \translation [1]{[#1]}%
\providecommand \BibitemOpen [0]{}%
\providecommand \bibitemStop [0]{}%
\providecommand \bibitemNoStop [0]{.\EOS\space}%
\providecommand \EOS [0]{\spacefactor3000\relax}%
\providecommand \BibitemShut  [1]{\csname bibitem#1\endcsname}%
\let\auto@bib@innerbib\@empty
\bibitem [{\citenamefont {{H. Kontani and S. Onari}}(2010)}]{Kontani_PRL104}%
  \BibitemOpen
  \bibfield  {author} {\bibinfo {author} {\bibnamefont {{H. Kontani and S.
  Onari}}},\ }\bibfield  {title} {\enquote {\bibinfo {title}
  {{Orbital-Fluctuation-Mediated Superconductivity in Iron Pnictides: Analysis
  of the Five-Orbital Hubbard-Holstein Model}},}\ }\href@noop {} {\bibfield
  {journal} {\bibinfo  {journal} {Phys. Rev. Lett.}\ }\textbf {\bibinfo
  {volume} {104}},\ \bibinfo {pages} {157001} (\bibinfo {year}
  {2010})}\BibitemShut {NoStop}%
\bibitem [{\citenamefont {{I. I. Mazin, D. J. Singh, M. D. Johannes and M. H.
  Du}}(2008)}]{MazinPRL2008}%
  \BibitemOpen
  \bibfield  {author} {\bibinfo {author} {\bibnamefont {{I. I. Mazin, D. J.
  Singh, M. D. Johannes and M. H. Du}}},\ }\bibfield  {title} {\enquote
  {\bibinfo {title} {{Unconventional Superconductivity with a Sign Reversal in
  the Order Parameter of LaFeAsO$_{1-x}$F$_x$}},}\ }\href@noop {} {\bibfield
  {journal} {\bibinfo  {journal} {Phys. Rev. Lett.}\ }\textbf {\bibinfo
  {volume} {101}},\ \bibinfo {pages} {057003} (\bibinfo {year}
  {2008})}\BibitemShut {NoStop}%
\bibitem [{\citenamefont {{K. Kuroki, S. Onari, R. Arita, H. Usui, Y. Tanaka,
  H. Kontani and H. Aoki}}(2008)}]{KurokiPRL101}%
  \BibitemOpen
  \bibfield  {author} {\bibinfo {author} {\bibnamefont {{K. Kuroki, S. Onari,
  R. Arita, H. Usui, Y. Tanaka, H. Kontani and H. Aoki}}},\ }\bibfield  {title}
  {\enquote {\bibinfo {title} {{Unconventional Pairing Originating from the
  Disconnected Fermi Surfaces of Superconducting LaFeAsO$_{1-x}$F$_x$}},}\
  }\href@noop {} {\bibfield  {journal} {\bibinfo  {journal} {Phys. Rev. Lett.}\
  }\textbf {\bibinfo {volume} {101}},\ \bibinfo {pages} {087004} (\bibinfo
  {year} {2008})}\BibitemShut {NoStop}%
\bibitem [{\citenamefont {{K. Seo, B. A. Bernevig and J.
  Hu}}(2008)}]{SeoPRL2008}%
  \BibitemOpen
  \bibfield  {author} {\bibinfo {author} {\bibnamefont {{K. Seo, B. A. Bernevig
  and J. Hu}}},\ }\bibfield  {title} {\enquote {\bibinfo {title} {{Pairing
  Symmetry in a Two-Orbital Exchange Coupling Model of Oxypnictides}},}\
  }\href@noop {} {\bibfield  {journal} {\bibinfo  {journal} {Phys. Rev. Lett.}\
  }\textbf {\bibinfo {volume} {101}},\ \bibinfo {pages} {206404} (\bibinfo
  {year} {2008})}\BibitemShut {NoStop}%
\bibitem [{\citenamefont {{H. Ding, P. Richard, K. Nakayama, K. Sugawara, T.
  Arakane, Y. Sekiba, A. Takayama, S. Souma, T. Sato, T. Takahashi, Z. Wang, X.
  Dai, Z. Fang, G. F. Chen, J. L. Luo and N. L. Wang}}(2008)}]{Ding_EPL}%
  \BibitemOpen
  \bibfield  {author} {\bibinfo {author} {\bibnamefont {{H. Ding, P. Richard,
  K. Nakayama, K. Sugawara, T. Arakane, Y. Sekiba, A. Takayama, S. Souma, T.
  Sato, T. Takahashi, Z. Wang, X. Dai, Z. Fang, G. F. Chen, J. L. Luo and N. L.
  Wang}}},\ }\bibfield  {title} {\enquote {\bibinfo {title} {{Observation of
  Fermi-surface-dependent nodeless superconducting gaps in
  Ba$_{0.6}$K$_{0.4}$Fe$_2$As$_2$}},}\ }\href@noop {} {\bibfield  {journal}
  {\bibinfo  {journal} {Europhys. Lett.}\ }\textbf {\bibinfo {volume} {83}},\
  \bibinfo {pages} {47001} (\bibinfo {year} {2008})}\BibitemShut {NoStop}%
\bibitem [{\citenamefont {{L. Zhao, H.-Y. Liu, W.-T. Zhang, J.-Q. Meng, X.-W.
  Jia, G.-D. Liu, X.-Li Dong, G.-F. Chen, J.-L. Luo, N.-L. Wang, W. Lu, G.-L.
  Wang, Y. Zhou, Y. Zhu, X.-Y. Wang, Z.-Y. Xu, C.-T. Chen and X.-J.
  Zhou}}(2008)}]{L_Zhao_CPL25}%
  \BibitemOpen
  \bibfield  {author} {\bibinfo {author} {\bibnamefont {{L. Zhao, H.-Y. Liu,
  W.-T. Zhang, J.-Q. Meng, X.-W. Jia, G.-D. Liu, X.-Li Dong, G.-F. Chen, J.-L.
  Luo, N.-L. Wang, W. Lu, G.-L. Wang, Y. Zhou, Y. Zhu, X.-Y. Wang, Z.-Y. Xu,
  C.-T. Chen and X.-J. Zhou}}},\ }\bibfield  {title} {\enquote {\bibinfo
  {title} {{Multiple Nodeless Superconducting Gaps in
  (Ba$_{0.6}$K$_{0.4}$)Fe$_2$As$_2$ Superconductor from Angle-Resolved
  Photoemission Spectroscopy}},}\ }\href@noop {} {\bibfield  {journal}
  {\bibinfo  {journal} {Chin. Phys. Lett.}\ }\textbf {\bibinfo {volume} {25}},\
  \bibinfo {pages} {4402} (\bibinfo {year} {2008})}\BibitemShut {NoStop}%
\bibitem [{\citenamefont {{L. Wray, D. Qian, D. Hsieh, Y. Xia, L. Li, J. G.
  Checkelsky, A. Pasupathy, K. K. Gomes, C. V. Parker, A. V. Fedorov, G. F.
  Chen, J. L. Luo, A. Yazdani, N. P. Ong, N. L. Wang, and M. Z.
  Hasan}}(2008)}]{Wray_PRB2008}%
  \BibitemOpen
  \bibfield  {author} {\bibinfo {author} {\bibnamefont {{L. Wray, D. Qian, D.
  Hsieh, Y. Xia, L. Li, J. G. Checkelsky, A. Pasupathy, K. K. Gomes, C. V.
  Parker, A. V. Fedorov, G. F. Chen, J. L. Luo, A. Yazdani, N. P. Ong, N. L.
  Wang, and M. Z. Hasan}}},\ }\bibfield  {title} {\enquote {\bibinfo {title}
  {{Momentum dependence of superconducting gap, strong-coupling dispersion
  kink, and tightly bound Cooper pairs in the high-$T_c$ (Sr, Ba)$_{1-x}$(K,
  Na)$_x$Fe$_2$As$_2$ superconductors}},}\ }\href@noop {} {\bibfield  {journal}
  {\bibinfo  {journal} {Phys. Rev. B}\ }\textbf {\bibinfo {volume} {78}},\
  \bibinfo {pages} {184508} (\bibinfo {year} {2008})}\BibitemShut {NoStop}%
\bibitem [{\citenamefont {{D. V. Evtushinsky, V. B. Zabolotnyy, T. K. Kim, A.
  A. Kordyuk, A. N. Yaresko, J. Maletz, S. Aswartham, S. Wurmehl, A. V. Boris,
  D. L. Sun, C. T. Lin, B. Shen, H. H. Wen, A. Varykhalov, R. Follath, B.
  B\"{u}chner and S. V. Borisenko}}()}]{Evtushinsky_PRB89}%
  \BibitemOpen
  \bibfield  {author} {\bibinfo {author} {\bibnamefont {{D. V. Evtushinsky, V.
  B. Zabolotnyy, T. K. Kim, A. A. Kordyuk, A. N. Yaresko, J. Maletz, S.
  Aswartham, S. Wurmehl, A. V. Boris, D. L. Sun, C. T. Lin, B. Shen, H. H. Wen,
  A. Varykhalov, R. Follath, B. B\"{u}chner and S. V. Borisenko}}},\ }\bibfield
   {title} {\enquote {\bibinfo {title} {{Strong electron pairing at the iron
  3$d_{xz,yz}$ orbitals in hole-doped BaFe$_2$As$_2$ superconductors revealed
  by angle-resolved photoemission spectroscopy}},}\ }\href@noop {} {\bibfield
  {journal} {\bibinfo  {journal} {Phys. Rev. B}\ }\textbf {\bibinfo {volume}
  {89}},\ \bibinfo {pages} {064514} (\bibinfo {year} {2014})}\BibitemShut
  {NoStop}%
\bibitem [{\citenamefont {{T. Shimojima, F. Sakaguchi, K. Ishizaka, Y. Ishida,
  T. Kiss, M. Okawa, T. Togashi, C.-T. Chen, S. Watanabe, M. Arita, K. Shimada,
  H. Namatame, M. Taniguchi, K. Ohgushi, S. Kasahara, T. Terashima, T.
  Shibauchi \emph{et al.}}}(2011)}]{Shimojima_Science332}%
  \BibitemOpen
  \bibfield  {author} {\bibinfo {author} {\bibnamefont {{T. Shimojima, F.
  Sakaguchi, K. Ishizaka, Y. Ishida, T. Kiss, M. Okawa, T. Togashi, C.-T. Chen,
  S. Watanabe, M. Arita, K. Shimada, H. Namatame, M. Taniguchi, K. Ohgushi, S.
  Kasahara, T. Terashima, T. Shibauchi \emph{et al.}}}},\ }\bibfield  {title}
  {\enquote {\bibinfo {title} {{Orbital-Independent Superconducting Gaps in
  Iron Pnictides}},}\ }\href@noop {} {\bibfield  {journal} {\bibinfo  {journal}
  {Science}\ }\textbf {\bibinfo {volume} {332}},\ \bibinfo {pages} {564}
  (\bibinfo {year} {2011})}\BibitemShut {NoStop}%
\bibitem [{\citenamefont {{G. F. Chen, Z. Li, J. Dong, G. Li, W. Z. Hu, X. D.
  Zhang, X. H. Song, P. Zheng, N. L. Wang and J. L.
  Luo}}(2008)}]{GF_ChenPRB78}%
  \BibitemOpen
  \bibfield  {author} {\bibinfo {author} {\bibnamefont {{G. F. Chen, Z. Li, J.
  Dong, G. Li, W. Z. Hu, X. D. Zhang, X. H. Song, P. Zheng, N. L. Wang and J.
  L. Luo}}},\ }\bibfield  {title} {\enquote {\bibinfo {title} {{Transport and
  anisotropy in single-crystalline SrFe$_2$As$_2$ and
  A$_{0.6}$K$_{0.4}$Fe$_2$As$_2$ (A = Sr, Ba) superconductors}},}\ }\href@noop
  {} {\bibfield  {journal} {\bibinfo  {journal} {Phys. Rev. B}\ }\textbf
  {\bibinfo {volume} {78}},\ \bibinfo {pages} {224512} (\bibinfo {year}
  {2008})}\BibitemShut {NoStop}%
\bibitem [{\citenamefont {{X.-P. Wang, P. Richard, Y.-B. Huang, H. Miao, L.
  Cevey, N. Xu, Y.-J. Sun, T. Qian, Y.-M. Xu, M. Shi, J.-P. Hu, X. Dai and H.
  Ding}}(2012)}]{XP_WangPRB85}%
  \BibitemOpen
  \bibfield  {author} {\bibinfo {author} {\bibnamefont {{X.-P. Wang, P.
  Richard, Y.-B. Huang, H. Miao, L. Cevey, N. Xu, Y.-J. Sun, T. Qian, Y.-M. Xu,
  M. Shi, J.-P. Hu, X. Dai and H. Ding}}},\ }\bibfield  {title} {\enquote
  {\bibinfo {title} {{Orbital characters determined from Fermi surface
  intensity patterns using angle-resolved photoemission spectroscopy}},}\
  }\href@noop {} {\bibfield  {journal} {\bibinfo  {journal} {Phys. Rev. B}\
  }\textbf {\bibinfo {volume} {85}},\ \bibinfo {pages} {214518} (\bibinfo
  {year} {2012})}\BibitemShut {NoStop}%
\bibitem [{\citenamefont {{A. V. Balatsky, I. Vekhter and J.-X.
  Zhu}}(2006)}]{BalatskyRMP}%
  \BibitemOpen
  \bibfield  {author} {\bibinfo {author} {\bibnamefont {{A. V. Balatsky, I.
  Vekhter and J.-X. Zhu}}},\ }\bibfield  {title} {\enquote {\bibinfo {title}
  {{Impurity-induced states in conventional and unconventional
  superconductors}},}\ }\href@noop {} {\bibfield  {journal} {\bibinfo
  {journal} {Rev. Mod. Phys.}\ }\textbf {\bibinfo {volume} {78}},\ \bibinfo
  {pages} {373} (\bibinfo {year} {2006})}\BibitemShut {NoStop}%
\bibitem [{\citenamefont {{P. Richard, T. Sato, S. Souma, K. Nakayama, H. W.
  Liu, K. Iwaya, T. Hitosugi, H. Aida, H. Ding and T.
  Takahashi}}(2012)}]{RichardAPL2012}%
  \BibitemOpen
  \bibfield  {author} {\bibinfo {author} {\bibnamefont {{P. Richard, T. Sato,
  S. Souma, K. Nakayama, H. W. Liu, K. Iwaya, T. Hitosugi, H. Aida, H. Ding and
  T. Takahashi}}},\ }\bibfield  {title} {\enquote {\bibinfo {title}
  {{Observation of momentum space semi-localization in Si-doped
  $\beta$-Ga$_2$O$_3$}},}\ }\href@noop {} {\bibfield  {journal} {\bibinfo
  {journal} {Appl. Phys. Lett.}\ }\textbf {\bibinfo {volume} {101}},\ \bibinfo
  {pages} {232105} (\bibinfo {year} {2012})}\BibitemShut {NoStop}%
\bibitem [{\citenamefont {{K. Iwaya, R. Shimizu, H. Aida, T. Hashizume and T.
  Hitosugi}}(2011)}]{IwayaAPL2011}%
  \BibitemOpen
  \bibfield  {author} {\bibinfo {author} {\bibnamefont {{K. Iwaya, R. Shimizu,
  H. Aida, T. Hashizume and T. Hitosugi}}},\ }\bibfield  {title} {\enquote
  {\bibinfo {title} {{Atomically resolved silicon donor states of
  $\beta$-Ga$_2$O$_3$}},}\ }\href@noop {} {\bibfield  {journal} {\bibinfo
  {journal} {Appl. Phys. Lett.}\ }\textbf {\bibinfo {volume} {98}},\ \bibinfo
  {pages} {142116} (\bibinfo {year} {2011})}\BibitemShut {NoStop}%
\bibitem [{\citenamefont {{M. M. Parish, J. Hu and B. A.
  Bernevig}}(2008)}]{ParishPRB78}%
  \BibitemOpen
  \bibfield  {author} {\bibinfo {author} {\bibnamefont {{M. M. Parish, J. Hu
  and B. A. Bernevig}}},\ }\bibfield  {title} {\enquote {\bibinfo {title}
  {{Experimental consequences of the s-wave $\cos(k_x)\cos(k_y)$
  superconductivity in the iron pnictides}},}\ }\href@noop {} {\bibfield
  {journal} {\bibinfo  {journal} {Phys. Rev. B}\ }\textbf {\bibinfo {volume}
  {78}},\ \bibinfo {pages} {144514} (\bibinfo {year} {2008})}\BibitemShut
  {NoStop}%
\bibitem [{\citenamefont {{D. Parker, O. V. Dolgov, M. M. Korshunov, A. A.
  Golubov and I. I. Mazin}}(2008)}]{ParkerPRB78}%
  \BibitemOpen
  \bibfield  {author} {\bibinfo {author} {\bibnamefont {{D. Parker, O. V.
  Dolgov, M. M. Korshunov, A. A. Golubov and I. I. Mazin}}},\ }\bibfield
  {title} {\enquote {\bibinfo {title} {{Extended $s_{\pm}$ scenario for the
  nuclear spin-lattice relaxation rate in superconducting pnictides}},}\
  }\href@noop {} {\bibfield  {journal} {\bibinfo  {journal} {Phys. Rev. B}\
  }\textbf {\bibinfo {volume} {78}},\ \bibinfo {pages} {134524} (\bibinfo
  {year} {2008})}\BibitemShut {NoStop}%
\bibitem [{\citenamefont {{A. V. Chubukov, D. V. Efremov and I.
  Eremin}}(2008)}]{ChubukovPRB78}%
  \BibitemOpen
  \bibfield  {author} {\bibinfo {author} {\bibnamefont {{A. V. Chubukov, D. V.
  Efremov and I. Eremin}}},\ }\bibfield  {title} {\enquote {\bibinfo {title}
  {{Magnetism, superconductivity, and pairing symmetry in iron-based
  superconductors}},}\ }\href@noop {} {\bibfield  {journal} {\bibinfo
  {journal} {Phys. Rev. B}\ }\textbf {\bibinfo {volume} {78}},\ \bibinfo
  {pages} {134512} (\bibinfo {year} {2008})}\BibitemShut {NoStop}%
\bibitem [{\citenamefont {{Y. Bang and H.-Y. Choi}}(2008)}]{BangPRB78}%
  \BibitemOpen
  \bibfield  {author} {\bibinfo {author} {\bibnamefont {{Y. Bang and H.-Y.
  Choi}}},\ }\bibfield  {title} {\enquote {\bibinfo {title} {{Possible pairing
  states of the Fe-based superconductors}},}\ }\href@noop {} {\bibfield
  {journal} {\bibinfo  {journal} {Phys. Rev. B}\ }\textbf {\bibinfo {volume}
  {78}},\ \bibinfo {pages} {134523} (\bibinfo {year} {2008})}\BibitemShut
  {NoStop}%
\bibitem [{\citenamefont {{Y. Wang, A. Kreisel, P. J. Hirschfeld and V.
  Mishra}}(2013)}]{Y_WangPRB87}%
  \BibitemOpen
  \bibfield  {author} {\bibinfo {author} {\bibnamefont {{Y. Wang, A. Kreisel,
  P. J. Hirschfeld and V. Mishra}}},\ }\bibfield  {title} {\enquote {\bibinfo
  {title} {{Using controlled disorder to distinguish s$_{\pm}$ and s$_{++}$ gap
  structure in Fe-based superconductors}},}\ }\href@noop {} {\bibfield
  {journal} {\bibinfo  {journal} {Phys. Rev. B}\ }\textbf {\bibinfo {volume}
  {87}},\ \bibinfo {pages} {094504} (\bibinfo {year} {2013})}\BibitemShut
  {NoStop}%
\bibitem [{\citenamefont {{P. W. Anderson}}(1959)}]{PW_AndersonJPCS11}%
  \BibitemOpen
  \bibfield  {author} {\bibinfo {author} {\bibnamefont {{P. W. Anderson}}},\
  }\bibfield  {title} {\enquote {\bibinfo {title} {{Theory of dirty
  superconductors}},}\ }\href@noop {} {\bibfield  {journal} {\bibinfo
  {journal} {J. Phys. Chem. Solids}\ }\textbf {\bibinfo {volume} {11}},\
  \bibinfo {pages} {26} (\bibinfo {year} {1959})}\BibitemShut {NoStop}%
\bibitem [{\citenamefont {{M. N. Gastiasoro, P. J. Hirschfeld and Brian M.
  Andersen}}(2014)}]{Gastiasoro_PRB89}%
  \BibitemOpen
  \bibfield  {author} {\bibinfo {author} {\bibnamefont {{M. N. Gastiasoro, P.
  J. Hirschfeld and Brian M. Andersen}}},\ }\bibfield  {title} {\enquote
  {\bibinfo {title} {{Origin of electronic dimers in the spin-density wave
  phase of Fe-based superconductors}},}\ }\href@noop {} {\bibfield  {journal}
  {\bibinfo  {journal} {Phys. Rev. B}\ }\textbf {\bibinfo {volume} {89}},\
  \bibinfo {pages} {100502} (\bibinfo {year} {2014})}\BibitemShut {NoStop}%
\bibitem [{\citenamefont {{W. Malaeb, T. Shimojima, Y. Ishida, K. Okazaki, Y.
  Ota, K. Ohgushi, K. Kihou, T. Saito, C. H. Lee, S. Ishida, M. Nakajima, S.
  Uchida, H. Fukazawa, Y. Kohori, A. Iyo, H. Eisaki, C.-T. Chen, S. Watanabe,
  H. Ikeda, and S. Shin}}(2012)}]{Malaeb_PRB86}%
  \BibitemOpen
  \bibfield  {author} {\bibinfo {author} {\bibnamefont {{W. Malaeb, T.
  Shimojima, Y. Ishida, K. Okazaki, Y. Ota, K. Ohgushi, K. Kihou, T. Saito, C.
  H. Lee, S. Ishida, M. Nakajima, S. Uchida, H. Fukazawa, Y. Kohori, A. Iyo, H.
  Eisaki, C.-T. Chen, S. Watanabe, H. Ikeda, and S. Shin}}},\ }\bibfield
  {title} {\enquote {\bibinfo {title} {{Abrupt change in the energy gap of
  superconducting Ba$_{1-x}$K$_x$Fe$_2$As$_2$ single crystals with hole
  doping}},}\ }\href@noop {} {\bibfield  {journal} {\bibinfo  {journal} {Phys.
  Rev. B}\ }\textbf {\bibinfo {volume} {86}},\ \bibinfo {pages} {165117}
  (\bibinfo {year} {2012})}\BibitemShut {NoStop}%
\bibitem [{\citenamefont {{D. C. Johnston}}(2010)}]{JohnstonAdv_Phys2010}%
  \BibitemOpen
  \bibfield  {author} {\bibinfo {author} {\bibnamefont {{D. C. Johnston}}},\
  }\bibfield  {title} {\enquote {\bibinfo {title} {{The Puzzle of High
  Temperature Superconductivity in Layered Iron Pnictides and
  Chalcogenides}},}\ }\href@noop {} {\bibfield  {journal} {\bibinfo  {journal}
  {Adv. Phys.}\ }\textbf {\bibinfo {volume} {59}},\ \bibinfo {pages} {803}
  (\bibinfo {year} {2010})}\BibitemShut {NoStop}%
\bibitem [{\citenamefont {{H. Ding, K. Nakayama, P. Richard, S. Souma, T. Sato,
  T. Takahashi, M. Neupane, Y.-M. Xu, Z.-H. Pan, A. V. Fedorov, Z. Wang, X.
  Dai, Z. Fang, G. F. Chen, J. L. Luo and N. L. Wang}}(2011)}]{Ding_JPCM2011}%
  \BibitemOpen
  \bibfield  {author} {\bibinfo {author} {\bibnamefont {{H. Ding, K. Nakayama,
  P. Richard, S. Souma, T. Sato, T. Takahashi, M. Neupane, Y.-M. Xu, Z.-H. Pan,
  A. V. Fedorov, Z. Wang, X. Dai, Z. Fang, G. F. Chen, J. L. Luo and N. L.
  Wang}}},\ }\bibfield  {title} {\enquote {\bibinfo {title} {{Electronic
  structure of optimally doped pnictide Ba$_{0.6}$K$_{0.4}$Fe$_2$As$_2$: a
  comprehensive angle-resolved photoemission spectroscopy investigation}},}\
  }\href@noop {} {\bibfield  {journal} {\bibinfo  {journal} {J. Phys: Condens.
  Matter}\ }\textbf {\bibinfo {volume} {23}},\ \bibinfo {pages} {135701}
  (\bibinfo {year} {2011})}\BibitemShut {NoStop}%
\bibitem [{\citenamefont {{K. Umezawa, Y. Li, H. Miao, K. Nakayama, Z.-H. Liu,
  P. Richard, T. Sato, J. B. He, D.-M. Wang, G. F. Chen, H. Ding, T. Takahashi
  and S.-C. Wang}}(2012)}]{UmezawaPRL2012}%
  \BibitemOpen
  \bibfield  {author} {\bibinfo {author} {\bibnamefont {{K. Umezawa, Y. Li, H.
  Miao, K. Nakayama, Z.-H. Liu, P. Richard, T. Sato, J. B. He, D.-M. Wang, G.
  F. Chen, H. Ding, T. Takahashi and S.-C. Wang}}},\ }\bibfield  {title}
  {\enquote {\bibinfo {title} {{Unconventional Anisotropic s-Wave
  Superconducting Gaps of the LiFeAs Iron-Pnictide Superconductor}},}\
  }\href@noop {} {\bibfield  {journal} {\bibinfo  {journal} {Phys. Rev. Lett.}\
  }\textbf {\bibinfo {volume} {108}},\ \bibinfo {pages} {037002} (\bibinfo
  {year} {2012})}\BibitemShut {NoStop}%
\bibitem [{\citenamefont {{S. V. Borisenko, V. B. Zabolotnyy, A. A. Kordyuk, D.
  V. Evtushinsky, T. K. Kim, I. V. Morozov, R. Follath and B.
  B\"{u}chner}}(2012)}]{BorisenkoSymmetry4}%
  \BibitemOpen
  \bibfield  {author} {\bibinfo {author} {\bibnamefont {{S. V. Borisenko, V. B.
  Zabolotnyy, A. A. Kordyuk, D. V. Evtushinsky, T. K. Kim, I. V. Morozov, R.
  Follath and B. B\"{u}chner}}},\ }\bibfield  {title} {\enquote {\bibinfo
  {title} {{One-Sign Order Parameter in Iron Based Superconductor}},}\
  }\href@noop {} {\bibfield  {journal} {\bibinfo  {journal} {Symmetry}\
  }\textbf {\bibinfo {volume} {4}},\ \bibinfo {pages} {251} (\bibinfo {year}
  {2012})}\BibitemShut {NoStop}%
\bibitem [{\citenamefont {{X. Lu, C. Fang, W.-F. Tsai, Y. Jiang and J.
  Hu}}(2012)}]{X_LuPRB85}%
  \BibitemOpen
  \bibfield  {author} {\bibinfo {author} {\bibnamefont {{X. Lu, C. Fang, W.-F.
  Tsai, Y. Jiang and J. Hu}}},\ }\bibfield  {title} {\enquote {\bibinfo {title}
  {{$s$-wave superconductivity with orbital-dependent sign change in
  checkerboard models of iron-based superconductors}},}\ }\href@noop {}
  {\bibfield  {journal} {\bibinfo  {journal} {Phys. Rev. B}\ }\textbf {\bibinfo
  {volume} {85}},\ \bibinfo {pages} {054505} (\bibinfo {year}
  {2012})}\BibitemShut {NoStop}%
\bibitem [{\citenamefont {{Z. P. Yin, K. Haule and G.
  Kotliar}}(2013)}]{Yin_ZP_orbital2013}%
  \BibitemOpen
  \bibfield  {author} {\bibinfo {author} {\bibnamefont {{Z. P. Yin, K. Haule
  and G. Kotliar}}},\ }\bibfield  {title} {\enquote {\bibinfo {title} {{Spin
  dynamics and an orbital-antiphase pairing symmetry in iron-based
  superconductors}},}\ }\href@noop {} {\bibfield  {journal} {\bibinfo
  {journal} {arXiv:1311.1188v1}\ } (\bibinfo {year} {2013})}\BibitemShut
  {NoStop}%
\bibitem [{\citenamefont {{Y.-B. Huang, P. Richard, X.-P. Wang, T. Qian, and H.
  Ding}}(2012)}]{YB_HuangAIP_Adv2}%
  \BibitemOpen
  \bibfield  {author} {\bibinfo {author} {\bibnamefont {{Y.-B. Huang, P.
  Richard, X.-P. Wang, T. Qian, and H. Ding}}},\ }\bibfield  {title} {\enquote
  {\bibinfo {title} {{Angle-resolved photoemission studies of the
  superconducting gap symmetry in Fe-based superconductors}},}\ }\href@noop {}
  {\bibfield  {journal} {\bibinfo  {journal} {AIP Adv.}\ }\textbf {\bibinfo
  {volume} {2}},\ \bibinfo {pages} {041409} (\bibinfo {year}
  {2012})}\BibitemShut {NoStop}%
\bibitem [{\citenamefont {{J.-P. Hu}}(2013)}]{HuJP_PRX3}%
  \BibitemOpen
  \bibfield  {author} {\bibinfo {author} {\bibnamefont {{J.-P. Hu}}},\
  }\bibfield  {title} {\enquote {\bibinfo {title} {{Iron-Based Superconductors
  as Odd-Parity Superconductors}},}\ }\href@noop {} {\bibfield  {journal}
  {\bibinfo  {journal} {Phys. Rev. X}\ }\textbf {\bibinfo {volume} {3}},\
  \bibinfo {pages} {031004} (\bibinfo {year} {2013})}\BibitemShut {NoStop}%
\bibitem [{\citenamefont {{N. Hao and J. Hu}}(2014)}]{Hao_PRB89}%
  \BibitemOpen
  \bibfield  {author} {\bibinfo {author} {\bibnamefont {{N. Hao and J. Hu}}},\
  }\bibfield  {title} {\enquote {\bibinfo {title} {{Sign Change in the Odd
  Parity Superconducting State of Iron-Based Superconductors}},}\ }\href@noop
  {} {\bibfield  {journal} {\bibinfo  {journal} {Phys. Rev. B}\ }\textbf
  {\bibinfo {volume} {89}},\ \bibinfo {pages} {045144} (\bibinfo {year}
  {2014})}\BibitemShut {NoStop}%
\bibitem [{\citenamefont {{J.-P. Hu and N.-N. Hao}}(2012)}]{HuJP_PRX2012}%
  \BibitemOpen
  \bibfield  {author} {\bibinfo {author} {\bibnamefont {{J.-P. Hu and N.-N.
  Hao}}},\ }\bibfield  {title} {\enquote {\bibinfo {title} {{$S_4$ Symmetric
  Microscopic Model for Iron-Based Superconductors}},}\ }\href@noop {}
  {\bibfield  {journal} {\bibinfo  {journal} {Phys. Rev. X}\ }\textbf {\bibinfo
  {volume} {2}},\ \bibinfo {pages} {021009} (\bibinfo {year}
  {2012})}\BibitemShut {NoStop}%
\bibitem [{\citenamefont {{I. I. Mazin}}(2011)}]{MazinPRB84}%
  \BibitemOpen
  \bibfield  {author} {\bibinfo {author} {\bibnamefont {{I. I. Mazin}}},\
  }\bibfield  {title} {\enquote {\bibinfo {title} {{Symmetry analysis of
  possible superconducting states in K$_x$Fe$_y$Se$_2$ superconductors}},}\
  }\href@noop {} {\bibfield  {journal} {\bibinfo  {journal} {Phys. Rev. B}\
  }\textbf {\bibinfo {volume} {84}},\ \bibinfo {pages} {024529} (\bibinfo
  {year} {2011})}\BibitemShut {NoStop}%
\bibitem [{\citenamefont {{M. Khodas and A.V. Chubukov}}(2012)}]{KhodasPRL108}%
  \BibitemOpen
  \bibfield  {author} {\bibinfo {author} {\bibnamefont {{M. Khodas and A.V.
  Chubukov}}},\ }\bibfield  {title} {\enquote {\bibinfo {title} {{Interpocket
  Pairing and Gap Symmetry in Fe-Based Superconductors with Only Electron
  Pockets}},}\ }\href@noop {} {\bibfield  {journal} {\bibinfo  {journal} {Phys.
  Rev. Lett.}\ }\textbf {\bibinfo {volume} {108}},\ \bibinfo {pages} {247003}
  (\bibinfo {year} {2012})}\BibitemShut {NoStop}%
\end{thebibliography}%

\end{document}